# Color Routing via Cross-Polarized Detuned Plasmonic Nanoantennas in Large Area Metasurfaces

Matteo Barelli[1], Andrea Mazzanti[2], Maria Caterina Giordano[1], Giuseppe Della Valle[2,3,*], Francesco Buatier de Mongeot[1,*]

[1] Dipartimento di Fisica, Università di Genova, Via Dodecaneso 33, I-16146 Genova, Italy

[2] Dipartimento di Fisica, Politecnico di Milano, Piazza L. da Vinci 32, I-20133 Milano, Italy

[3] IFN-CNR, Piazza L. da Vinci 32, I-20133 Milano, Italy

* *To whom correspondence may be addressed:* giuseppe.dellavalle@polimi.it and buatier@fisica.unige.it

**TOC GRAPHIC**

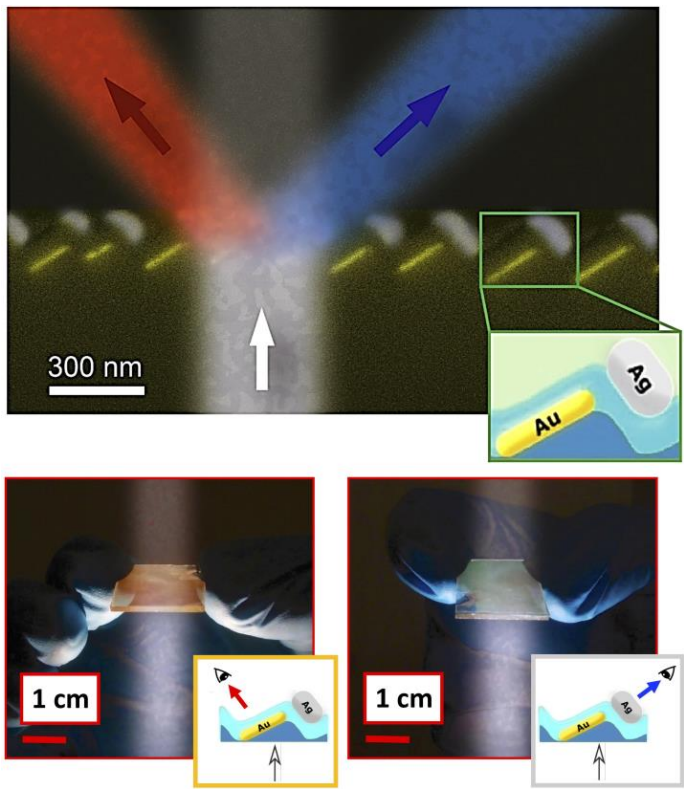

## ABSTRACT

Bidirectional nanoantennas are of key relevance for advanced functionalities to be implemented at the nanoscale, and in particular for color routing in an ultracompact flat-optics configuration. Here we demonstrate a novel approach avoiding complex collective geometries and/or restrictive morphological parameters, based on cross-polarized detuned plasmonic nanoantennas in a uniaxial (quasi-1D) bimetallic configuration. The nanofabrication of such a flat-optics system is controlled over a large-area (cm$^2$) by a novel self-organized technique exploiting ion-induced nanoscale wrinkling instability on glass templates to engineer tilted bimetallic nanostrip dimers. These nanoantennas feature broadband color routing with superior light scattering directivity figures, which are well described by numerical simulations and turn out to be competitive with the response of lithographic nanoantennas. These results demonstrate that our large-area self-organized metasurfaces can be implemented in real world applications of flat optics color routing from telecom photonics to optical nanosensing.

**INTRODUCTION**

The investigation of optical phenomena at the nanoscale has witnessed an amazing development in the last decades, with particular interest to metallic nanostructures, supporting so-called plasmonic resonances and subsequent spatial localization of intense optical fields far below the diffraction limit [1–4]. A key feature of subwavelength metallic structures is their capability to provide amplified light scattering and strong near field confinement at the resonant frequency, thus behaving as optical nanoantennas [5–10]. Plasmonic nanoantennas have been thoroughly developed to demonstrate challenging functionalities in flat-optics nanodevices [11,12,13], including waveplates [14,15], polarization splitters [16,17], directional nanoemitters [18–20], unidirectional antennas [21–23], and multi-directional color routers [24–27]. The availability of broadband and highly directive optical antennas is crucial for the development of a wide range of applications from optical sensing [2,28,29] to photon harvesting [12,30,31] and biosensing [32–35]. When dealing with color routing, i.e. wavelength selective multi-directional scattering, different strategies, comprising single or multi-element antenna systems, and different materials, have so far been explored: as summarized below, such approaches to color routing predominantly exploit complex design of the nanoantenna geometry, inter-antenna gap and chemical composition which require cumbersome top-down nanofabrication methods.

A first approach relies on multi-element nanoantennas. The simplest configuration exploits the scattering interference between two asymmetric nanoantenna elements whose resonant dipoles oscillate with a large phase shift with respect to the illuminating field, in the spectral range in-between their resonances [24,25,36,37]. The wavelength selectivity, enabling color routing operation, is obtained via a symmetry breaking induced, e.g., by using two different materials for the two nano-antenna elements [24,37]. The unidirectionality can be also enhanced with several elements in order to realize a downscaling of the classic RF Yagi-Uda antenna [21,38,39] or other array configurations [38,40,41]. Another approach exploits the near field interference of multiple resonances (e.g., electric and magnetic) simultaneously excited in the same antenna element [26,42–44]. Plasmonic nanoantennas can be engineered to achieve unidirectionality by resorting to specific antenna geometries [45,46], e.g. gold split-ring resonators [47] or even nanodisk antennas [48], but generally only dielectric and hybrid metal/dielectric antennas are suited for color routing due to the strength and phase shifts of magnetic modes in dielectric materials [26,49]. The third approach is based upon wavefront manipulation by highly ordered metasurfaces, with meta-atoms made of either single or multi-element nanoantennas, providing either wavelength independent unidirectional scattering [50], or wavelength selective narrow-band bi-directional color routing operation [27,51]. Very recently, active metasurfaces for bias controlled directional scattering have also started to gain attention [52]. All the outlined strategies require heavy computational effort and time-consuming top down nanofabrication processes to control inter-antennas spacing, shape and material composition, which are costly and inherently limited to small areas.

In this work we achieve broadband and highly effective color routing functionalities over large area (cm$^2$) with much more relaxed fabrication constraints and requirements compared to the state-of-the-art antenna design. Our idea is based on a novel flat-optics approach to bi-directional passive color routing at optical frequencies with cross-polarized detuned plasmonic nano-antennas in a uniaxial (quasi-1D) bimetallic configuration. The directional routing mechanism is based purely on the relative tilt and material composition of the nanoantenna elements, rather than on the interference of their resonant optical modes, which is actually inhibited by the cross-polarization configuration. We demonstrate that our strategy can be effectively implemented via self-organized nanofabrication based on controlled anisotropic nanoscale wrinkling in transparent templates and maskless confinement of plasmonic bi-directional antennas.

**RESULTS AND DISCUSSION**

The concept behind our approach is summarized in Fig. 1. Let's consider the two-dimensional (2D) scattering problem of two lines of cross-polarized electric dipolar scatterers (having out-of-plane continuous translational invariance), excited by a uniform plane wave with in-plane electric field (i.e. linear TM polarization) (Fig. 1a). If the two scatterers exhibit a resonant behavior with well separated scattering peaks (Fig. 1b), e.g., one in the blue ($\lambda_1$) and one in the red ($\lambda_2$), the blue light is mainly scattered at 90° with respect to the red light, with negligible superposition of the two spectral components in the far field along the two orthogonal directions.

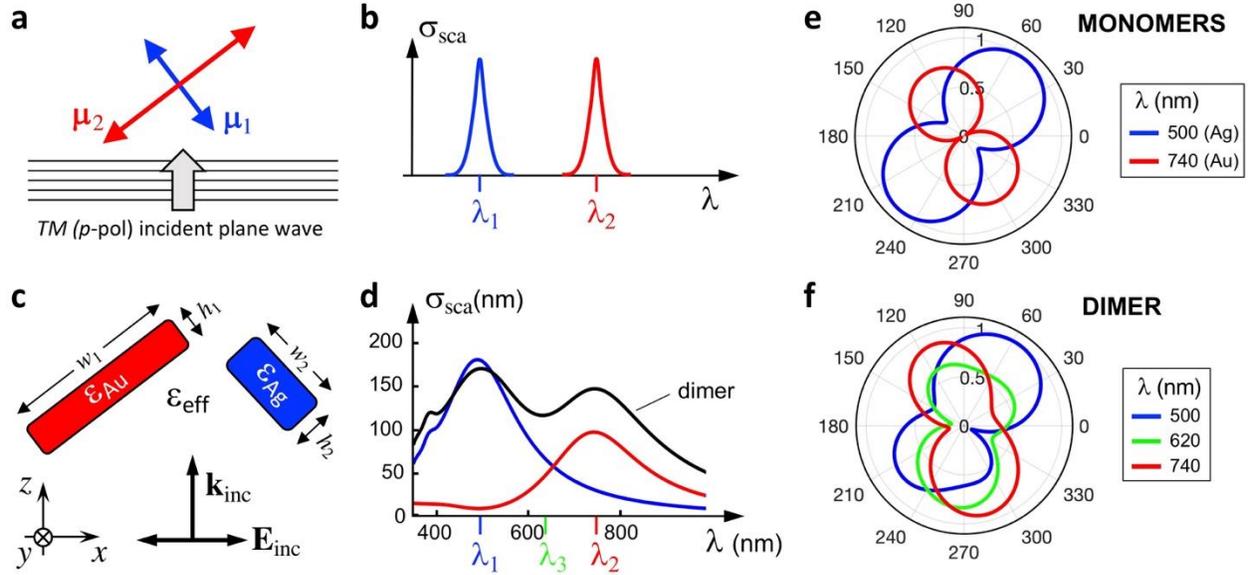

**Figure 1:** *(a) Sketch of the 2D cross-polarized electrical dipoles for color routing with (b) detuned scattering spectra; (c) implementation with bimetallic nanostrips (having continuous translational invariance along the out-of-plane y-axis) and (d) the actual scattering spectra of the gold monomer (red), of the silver monomer (blue) and of the dimer (black), retrieved by full-wave 2D numerical simulations; (e) scattering diagram of the nanostrip monomers at their peak resonance wavelengths; (f) same as (e) for the nanostrip dimer, also showing the scattering diagram at an intermediate wavelength (in-between the two resonances).*

Such an ideal system can be implemented with a couple of tilted metallic nanostrips, which are well known to behave as dipolar scatterers, having a pronounced resonance under TM light, enabled by a plasmonic response [53]. The latter can be controlled and tuned by acting on the width (*w*) and height (*h*) of the nanostrips, as well as on the metal permittivity. Basically, the fundamental dipolar resonance wavelength of the nanostrip linearly scales with *w*, and is inversely proportional to *h*. Also, note that for a given geometrical configuration, silver nanostrips resonate at shorter wavelengths compared to gold nanostrips [53]. The latter circumstance indicates that a hybrid Au/Ag configuration for the dimer should be more flexible to achieve detuned plasmonic resonances from the two nanostrips, which is a key feature for efficient color routing. Having this in mind, we designed a two dimensional plasmonic dimer made of a gold nanostrip with ($w_1,h_1$) = (100,13) nm tilted at about 40° with respect to the x-axis and a silver nanostrip with ($w_2,h_2$) = (*90,34*) nm tilted at about -50° (Fig. 1c). Finite element method (FEM) numerical analysis of the nanostrips scattering spectrum (Fig. 1d) retrieves a peak resonance wavelength at $\lambda_1$ = 500 nm for silver and at $\lambda_2$ =740 nm for gold isolated nanostrips (or monomers). The scattering diagram as a function of the polar angle $\Theta$ for the isolated monomers (Fig. 1e) exhibits the typical pattern of a dipolar scatterer, with two main lobes orthogonal to the major axis of the nanostrip. In the dimer configuration, the scattering diagram, even though being more complex because of the near-field coupling between the two monomers, preserves the key features of the individual scattering patterns at the two resonance wavelengths (Fig. 1f). In particular, note the wavelength dependent bidirectional behavior, characterized by a strong rejection of the blue wing of the spectrum at $\Theta$ ~ 140° and of the red wing at $\Theta$ ~ 40°, implementing the desired color routing functionality.

Even though conceptually simple, the design above detailed poses a fundamental challenge in terms of fabrication, because of the tilted configuration of the nanostrips and the requirement of a large area nanopatterning to allow flat-optics operation. Here, a self-organized nanofabrication method, based on anisotropic nanoscale wrinkling in glasses, enables the effective engineering of tilted plasmonic nanostrip antennas in the form of cross-polarized metallic dimers lying on opposite ridges of a faceted dielectric template.

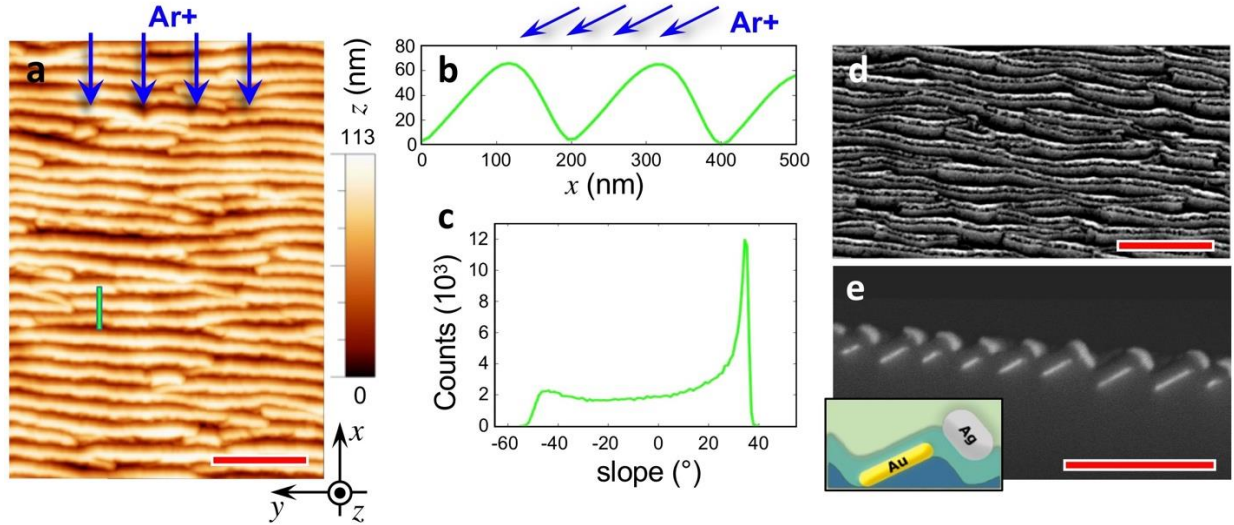

**Figure 2:** *(a) Atomic Force Microscopy (AFM) topography of the fabricated rippled glass template. Scale bar 1μm (b) AFM cross section of nanoripples corresponding to green line in panel (a). (c) Histogram of the slope distribution extracted by the AFM image of panel (a). (d) Scanning Electron Microscope (SEM) images of the bimetallic metasurface detected in top view and (e) cross section configuration. Scale bars 1μm and 600 nm, respectively. The inset represents a sketch of the unit cell cross section of the bimetallic surface.*

A low-cost glass substrate is irradiated with a defocused Ar$^+$ ion beam set at an incident angle of $\theta = 30°$ and at a low energy of 800 eV. The glass temperature is fixed at about 680 K during the Ion Beam Sputtering (IBS) process. A quasi 1-D rippled pattern is obtained all over the macroscopic sample surface, with a wavevector parallel to the ion beam direction (Fig. 2a). The glass ripples are elongated for a length of several micrometers and show a remarkable degree of long-range order. The nanostructures are characterized by a steep asymmetric saw-tooth profile with a pronounced vertical dynamic of approximately 90 nm (Fig. 2b) and a periodicity of about 200 nm (see the 2D self-correlation function of the AFM topography in Fig. S1). The ridges directly exposed to the ion beam develop a broad facet slope distribution (negative values in the slope frequency plot of Fig. 2c) with a characteristic slope peaked at about -50°. The opposite ridges (positive values in Fig. 2c) develop wider and very defined facets with slopes peaked at +35°.

As reported by some of the authors in a recent study [54], raising the substrate temperature near the glass transition is crucial to activate a peculiar solid-state wrinkling instability which strongly enhances the nanopattern growth, acting in parallel with the ion beam erosion process. Under this condition, high aspect ratio ordered 1D templates are obtained, with a strong improvement with respect to state-of-the-art large area IBS nanopattering of semiconductor and insulating substrates at room temperature, which commonly yields low-aspect ratio, disordered ripples [55,56].

Such high aspect-ratio 1D rippled templates, with well-defined tilted facets are the ideal platform for the maskless confinement of plasmonic uniaxial nano-strip antennas (NSA) with controlled tilt and morphology. The controlled growth is achieved by grazing angle physical vapor deposition of metal atoms exploiting shadowing effects on the tilted ripple facets [57]. In order to fabricate the bimetallic dipolar antennas depicted in the theoretical model of Fig. 1c, we first confined Au NSA by glancing evaporation on the wider rippled facets which are tilted at +35° (see inset sketch in Fig. 2e). By the statistical analysis of SEM cross section images (Fig. 2e) we measured the average Au NSA width $w \approx 105$ nm, controlled by the periodicity of the underlying template and by the local slope of the illuminated ripple facets. The thickness of the Au NSA reads $h \approx 12$ nm and is determined by the sublimated metal dose and deposition angle (see Supporting Information for details). As a second step, an insulating layer of sub-stoichiometric silica was grown on the sample at normal incidence by Radio Frequency Magnetron Sputtering. Finally, Ag NSA were confined on the narrower and steeper rippled facets with an average slope of -50°, on top of the conformally grown silica layer. Ag NSA average $w$ and $h$ are about 83 nm and 36 nm, respectively. The average dielectric gap separating the Au and Ag NSA corresponds to 40 nm, again evaluated from the SEM cross section analysis. This method enables the controlled growth of subwavelength bimetallic antennas whose length exceeds several micrometers and form a large area quasi 1-D metasurface, as demonstrated by the top-view and cross-section SEM images (Fig. 2d, 2e). The image of the sample cross section (Fig. 2e) evidences the selective lateral confinement of nanoantennas on top of the tilted ridges of the ripple, thus well mimicking the cross section of the ideal structure sketched in the inset of Fig. 2e.

Optical transmittance measurements at normal incidence were performed on the sample using a halogen/deuterium lamp source. Scattering measurements were instead acquired with a custom-made scatterometer set-up, following the scheme of Fig. 3l, m (see Supporting Information for details). The sample was illuminated from the glass side at normal incidence and the scattered light was collected at a fixed polar angle $\Theta = 50°$, as a function of the azimuthal angle $\Phi$. For both transmittance and scattering measurements, the light source was linearly polarized orthogonally to the NSA long axis (i.e. TM polarization). The same optical measurements were performed also on two other samples: Au NSA, and Ag NSA, fabricated with the same morphological and geometrical configuration as in the complete bimetallic Au-Ag NSA. This allows us to investigate the properties of the final system as well as the optical behavior of its two building blocks. The optical transmittance and the scattering intensities measured for the Au, Ag and Au-Ag NSA samples are shown in Fig. 3d-f and Fig. 3g-i, respectively. Remarkably, for all the three configurations a directional scattering maximum is detected, which is resonant to the localized surface plasmon mode (i.e. minimum in transmittance in Fig. 3d-f).

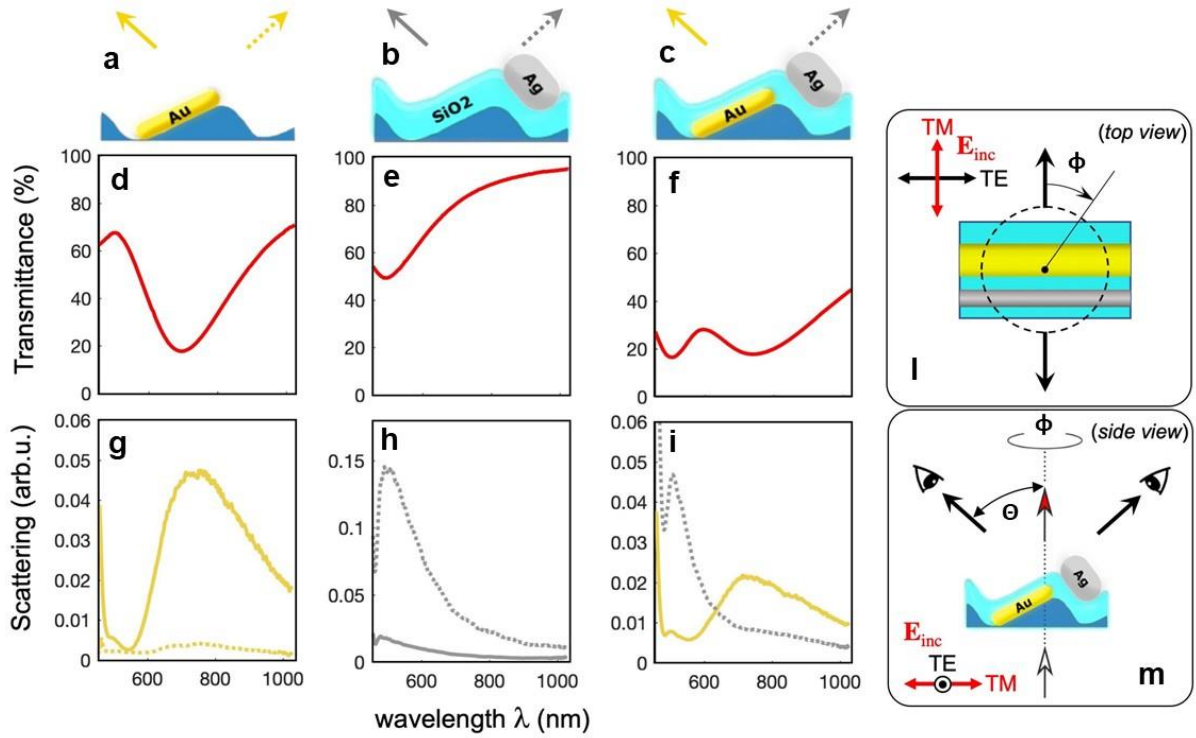

**Figure 3:** *(a,b,c) Sketch of the Au, Ag and Au/Ag NSA meta-atom cross section, respectively. (d,e,f) Measured optical transmittance at normal incidence for Au, Au and Au/Ag NSA arrays, respectively. All the spectra are normalized to the transmittance of the bare glass substrate. (g,h,i) Scattered light intensity detected at $\Theta = 50°$, $\Phi = 0°$ (solid curve) and at $\Theta = 50°$, $\Phi = 180°$ (dashed curve) for Au, Ag and Au/Ag NSA arrays, respectively. (l,m) Sketch of the optical set-up exploited for scattering measurements respectively shown in top- and side- view.*

As anticipated above, the scattering pattern of a plasmonic NSA is expected to resemble the one of a dipolar antenna. The transmitted scattering maximum is then expected to be centered along the direction normal to the NSA major axis. When $\Phi = 0°$, and $\Theta = 50°$, scattered light collection is essentially facing the +35° tilted Au NSA (as it can be clearly appreciated in the inset of Figs. 3l, m), i.e. the direction where the scattering intensity is expected to be stronger. Indeed, for this collection configuration (solid yellow curve in Fig. 3g), the measured scattering intensity shows an intense and broad maximum centered around 750 nm. Note that the scattering maximum and the plasmonic transmittance dip are resonant (the slight frequency shift being ascribable to the well-known mismatch between absorption and scattering resonances in plasmonic nanostructures), thus clearly demonstrating the plasmonic nature of the enhanced scattered light. When the collection of the scatterometer is set to the azimuthal angle $\Phi = 180°$, the scattered light is measured along the direction of the Au NSA major axis. As a consequence, the scattering intensity drops by about an order of magnitude, in accord with the radiation pattern simulations of Fig. 1e. Fig. 3e shows the transmittance measurement for the Ag NSA, with the Ag nanostrips confined on a silica layer grown upon the rippled glass template, to preserve the effective refractive index of the medium surrounding the NSA in the complete bimetallic NSA. The plasmonic transmittance dip is blue-shifted to 500 nm compared to the Au NSA, due to the different Ag NSA aspect ratio ($w/h$), and higher permittivity (in modulus) of silver compared to gold [58]. This time, when $\Phi = 180°$ (keeping $\Theta = 50°$) the scattered light collection is perfectly aligned to the normal of the Ag nanostrips, which are tilted at -50°, and an intense scattering peak is recorded at resonance, i.e. around 500 nm, near the edge of our laser source spectrum (dashed grey curve in Fig. 3h). Note that the resonant scattering intensity of the Ag NSA is about 3 times higher than in the Au NSA, mainly because of the better optimized collection angle with respect to the NSA axis and also thanks to the higher optical density of Ag, which implies a larger Ag scattering cross section [53]. When the azimuthal angle is set to $\Phi = 0°$ the scattered light is collected in a direction parallel to the Ag NSA axis, which corresponds to the direction of minimum scattered intensity emission pattern (cf. Fig. 1). As a consequence, the collected scattering intensity considerably drops down (solid grey curve in Fig. 3h). Finally, the transmittance spectrum for the complete bimetallic Ag-Au NSA is shown in Fig. 3f. Two separate transmittance dips are distinguishable at 500 nm and 740 nm, which are associated to the plasmonic resonances of the Ag NSA and Au NSA monomers, respectively. When the azimuthal angle is fixed to Φ = 0° the collection of the scattering intensity faces the Au NSA and a scattering maximum is recorded at the Au NSA resonance of 740 nm (solid yellow curve in Fig. 3g), with negligible contributions from the Ag NSA (cf. Fig 3h, solid grey curve). When the azimuthal angle is set to Φ = 180°, the collection direction faces the Ag NSA, and a scattering intensity peak is recorded at the Ag NSA resonance of 500 nm, with negligible contribution from the Au NSA (cf. Fig 3g, dashed yellow curve).

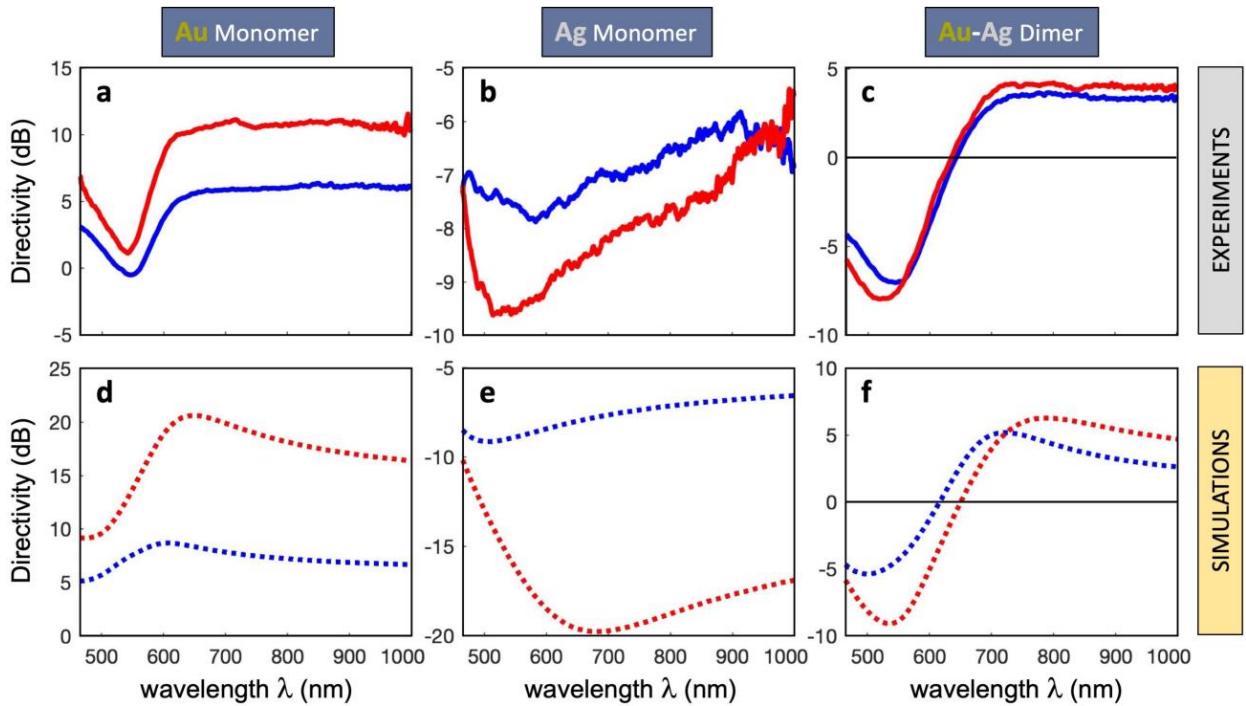

**Figure 4:** *Measured (a, b, c) and simulated (d, e, f) directivities at two different polar angles (Θ = 30° in blue, Θ = 50° in red) for: (a, d) Au NSA monomer, (b, e) Ag NSA monomer and (c, f) Au-Ag NSA dimer configurations, respectively.*

Note also that the scattering from the Au-Ag dimeric NSA is partially attenuated with respect to the case of individual Au or Ag NSA building blocks, mostly because of non-resonant shadowing effects. Actually, even though hybridization is negligible in our detuned cross-polarized configuration (in agreement with the numerical simulations of Fig. 1), each monomer is placed within the cross-section region of the other, thus causing an effective partial attenuation of the beam. Remarkably, the detected directivity figures highlight the high efficiency of self-organized bimetallic antennas in wavelength selective bidirectional scattering. This response is well explained in terms of the cross-polarized detuned plasmonic nanoantenna model. Provided the tilted nanoantennas configuration, the Ag and Au NSA monomer resonantly scatters light to the "right side" at about 500 nm wavelength, and to the "left side" at about 740 nm wavelegth, respectively. This color routing action taking place over the whole sample macroscopic area can be clearly appreciated in the digital camera images of Fig. S3.

To quantify the color routing performance in our samples, we plot in Fig. 4 the *Directivity spectrum* (*D*), defined as follows:

$$D(\lambda; \Theta) = 10 \times \log_{10}\left[\frac{S(\Phi = 0°, \Theta; \lambda)}{S(\Phi = 180°, \Theta; \lambda)}\right]$$

The Directivity spectra at polar angles $\Theta = 30°$ and $\Theta = 50°$ are shown in Fig. 4 (top panels), together with the corresponding numerical simulations (bottom panels). Note that the vertical axis dB scale is different for every panel of Fig. 4 for better reading of the traces. In particular, as can be appreciated in Fig. S4, the signal-to-noise ratio of the measured directivity data is indeed similar for all the three different considered NSAs configurations (Au monomer, Ag monomer and Au-Ag dimer). Note that the monomeric configurations based on Au (Fig. 4a) or Ag (Fig. 4b) NSAs, even though giving rise to a wavelength dependent directional scattering, are not capable of bidirectional functionality, having no sign changes in their Directivity. Conversely, the bimetallic Au-Ag NSA uniquely provides an inversion of *D* as a function of wavelength. The sign flips from negative to positive by increasing λ from the blue wing of the spectrum (dominated by the Ag NSA scattering) to the red one (dominated by Au NSA scattering), the zero being at around 620 nm (i.e. in correspondence to a local maximum of the transmittance spectrum of Fig. 3f, placed in-between the two orthogonal plasmonic resonances of the dimer configuration). Notably, the maximum directivity values for both the monomer and dimer configurations are well within the same order of magnitude reported for lithographic antennas and, at the same time, show a uniform broadband response which is particularly evident in the NIR range [24,26,45]. The quasi-1D nature of the arrays and the polydispersion of the NSA morphology lead to the absence of collective grating effects or surface lattice resonances and also ensure the spectrally broad response of the LSP resonance (cf. Ref. [59] for details and Fig. S1). Therefore, the optical properties of the NSAs are dominated by the resonant features of the unit cell, which is the typical regime of metasurface operation. Thus, the color routing properties of the system are not related to complex and/or restrictive collective geometries and morphological parameters, greatly relaxing the fabrication demands and issues.

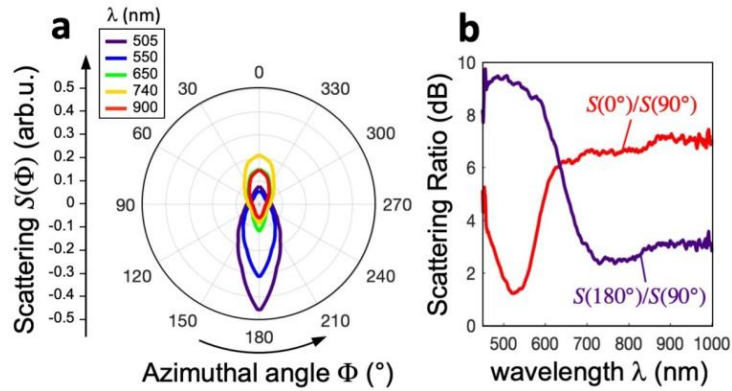

**Figure 5:** *(a) Azimuthal plot of the scattered light at polar angle $\Theta = 50°$ for 5 different wavelengths. (b) 90° scattering rejection ratio as a function of wavelength for left (red) and right (purple) routing operation.*

Finally, it is worth noting that even though our implementation of the NSAs employs finite length (i.e. few μm long) structures, thus breaking the continuous translational invariance of the ideal 2D configuration along *y*-axis (Fig. 2a), the measured azimuthal scattering pattern turns out to be well concentrated along the $\Phi = 0°$ and $\Phi = 180°$ directions of the color routing (Fig. 5a). This is especially true at around the plasmonic resonance wavelengths, as elucidated by the 90° scattering rejection ratio of Fig. 5b, defined as the ratio (in dB scale) between the light scattered at $\Phi = 0°$ (or $\Phi = 180°$) and that scattered at $\Phi = 90°$ for a given polar angle ($\Theta = 50°$ in the present case). The finite length of the nanostrips and their slight deviations to parallelism along *y*-axis (Fig. 2a), does not prevent to consider the system as substantially invariant in the direction of the ridges, which is consistent with our design and modeling strategy. A related issue pertains to the polarization sensitivity of our structures, which belongs to the quasi-1D nature of the array configuration [57,60]. To address this point, we have repeated all the optical experiments with TE polarization (electric field of the source parallel to the nanostrips i.e. orthogonal to their cross-sectional plane – see top-view sketch in Figure 3l). The results indicate a very shallow directional scattering in terms of the azimuthal angle, and the absence of bidirectional wavelength sensitivity which is consistent with the basic theoretical argument according to which no plasmonic resonances can be excited under TE illumination in a two-dimensional configuration.

**CONCLUSIONS**

We demonstrated broadband color routing from cross-polarized detuned plasmonic nanoantennas, fabricated over a large (cm$^2$) area via self-organized metal confinement on a faceted glass template. The latter represents a natural platform for guiding the growth of vertically tilted nanoantenna arrays in a single maskless step. The resulting plasmonic metasurfaces exhibit highly directional and wavelength selective properties which are widely tunable and not related to complex and/or restrictive collective geometries and morphological parameters, greatly relaxing the fabrication demands and related critical issues. All the main optical and scattering features of these structures can be effectively predicted by numerical simulations in which the key parameters (such as antennas tilt and materials dielectric function) can be readily implemented by our self-organization methods. The scattering directivities of the proposed large-area beam-splitting metasurfaces are competitive with the figures of merit of lithographically patterned nanoantennas. We thus believe that our results can open the way to the use of flat-optics broadband color routers in large area applications, with potential impact on a wide range of real world devices involving e.g. dichroic beam splitters, broadband polarizers and multiplexed plasmonic biosensors.

## ASSOCIATED CONTENT

Supporting Information contains additional details and a description of the experimental methods employed. The document is organized in the following sections: Fabrication of Tilted Optical Nanoantennas, Morphological Characterization, Numerical Optical Model, Optical Characterization.

## AUTHOR INFORMATION

**Corresponding authors**

Correspondence to Giuseppe Della Valle or Francesco Buatier De Mongeot.

**Contributions**

M.B., F.B.d.M. and G.D.V. conceived the idea. M.B., M.C.G. and F.B.d.M. performed the experiments and conceived the experimental setup. A.M. performed the electromagnetic simulations. All the authors contributed to the writing of the manuscript. F.B.d.M. and G.D.V. also supervised the project.

## ACKNOWLEDGMENTS

We acknowledge the financial support from the Italian Ministry of Education, University and Research (MIUR), through the PRIN 2015 Grant No. 2015WTW7J3.

# Supporting Information

# Color Routing via Cross-Polarized Detuned Plasmonic Nanoantennas in Large Area Metasurfaces

Matteo Barelli[1], Andrea Mazzanti[2], Maria Caterina Giordano[1], Giuseppe Della Valle[2,3,*], Francesco Buatier de Mongeot[1,*]

[1] Dipartimento di Fisica, Università di Genova, Via Dodecaneso 33, I-16146 Genova, Italy

[2] Dipartimento di Fisica, Politecnico di Milano, Piazza L. da Vinci 32, I-20133 Milano, Italy

[3] IFN-CNR, Piazza L. da Vinci 32, I-20133 Milano, Italy

* *To whom correspondence may be addressed:*

giuseppe.dellavalle@polimi.it and buatier@fisica.unige.it

**TILTED OPTICAL NANOANTENNAS FABRICATION**

A soda-lime glass substrate (20 x 20 x 2 mm) is repeatedly rinsed in ethanol and acetone. The sample is then placed in a custom-made vacuum chamber and irradiated with an 800 eV low energy defocused $Ar^+$ ion beam (gas purity N5.0). A biased tungsten filament avoids charge build-up through thermionic electron emission. The ion beam illuminates the glass surface at an incident angle of $\theta = 30°$ with respect to its normal. The ion fluence corresponds to $1.4 \times 10^{19}$ ions/cm$^2$ and the glass temperature is fixed at about 680 K during the Ion Beam Sputtering (IBS) process. After the rippled pattern is formed on the glass surface, thermal Au deposition is performed on the rippled facets at a glancing angle $\theta = 55°$ with respect to the flat sample normal. The Au beam directly illuminates the glass facets tilted at +35° while the opposite facets are completely shadowed. By means of a calibrated quartz microbalance, the thickness $h$ of the Au stripes can be evaluated by basic geometrical arguments given the Au thickness ($h_0$) deposited on a flat surface facing the crucible at normal incidence and the average slope of the illuminated facet measured with AFM as $h = h_0 \times cos(55°- 35°)$. The sample is then put in a custom-made RF sputtering chamber where a layer of $SiO_2$ is conformally grown all over the surface using a 2" fused silica target. The silica layer thickness was monitored by means of a calibrated quartz microbalance. The RF sputtering experiment is run in Argon atmosphere at a power $P = 60$ $W$, sample-target distance $d = 8.5$ $cm$ and total pressure of about $P = 7 \times 10^{-2}$ $mbar$. Finally, Ag stripes are confined on the rippled facets tilted at -50°, now coated with a conformal $SiO_x$ layer, by using the same strategy and arguments already described for the Au ones.

**MORPHOLOGICAL CHARACTERIZATION**

The rippled glass template morphology was characterized by means of an atomic force microscope (Nanosurf S Mobile), running in tapping mode. The average periodicity and slope of the glass ripples were computed from the statistical analysis of AFM topographies by means of WSxM software (Fig. SI1). Top view and cross section back scattered electrons images of the bimetallic nanoantennas array were acquired by means of a scanning electron microscope (Hitachi VP-SEM SU3500), operating in the 10-15 kV accelarating voltage range. Statistical analysis was performed on the SEM images by means of ImageJ software in order to evaluate the average antennas width and average antennas and silica gap thicknesses.

Fig. S1a shows a typical AFM topography of the self-organized nanostructured glass template. The template average periodicity $\lambda$ is estimated from the real space distance between the maximum and the secondary neighboring peaks in the 2D self-correlation function (Figs. S1b and S1c).

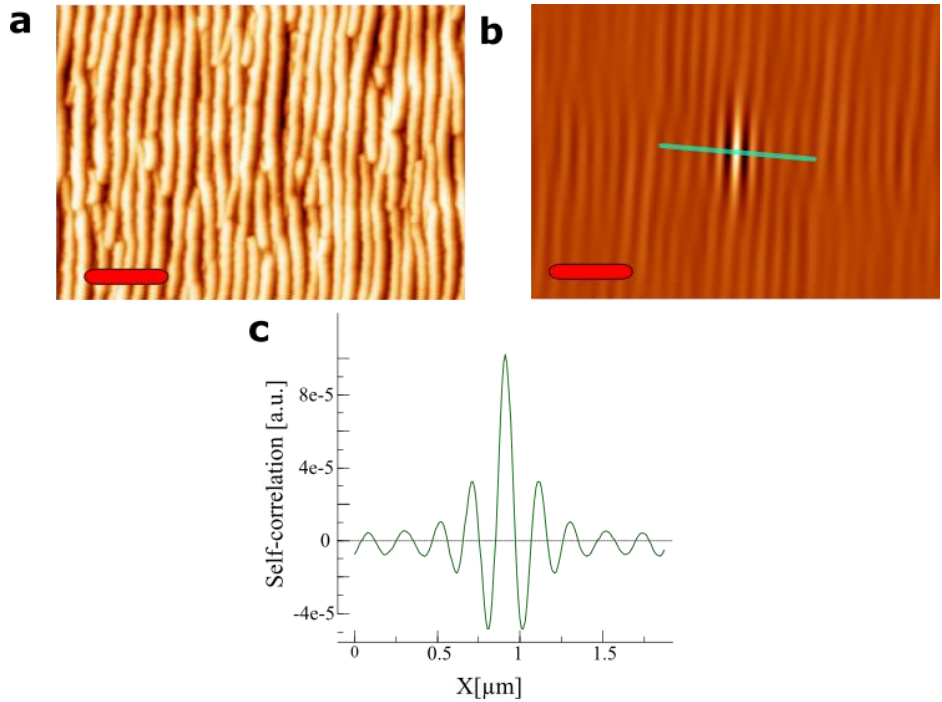

**Figure S1:** *(a) AFM topography, (b) self-correlation function, (c) line profile (bottom panel) of the self-correlation function, extracted in correspondence of the green line in (b). The red scale bars correspond to 1 µm.*

It's worth to note how the 2D self-correlation function of the rippled patterns rapidly decays to negligible values away from the central maxima, as the pattern loses its morphological coherence within 2-3 unit cells. This prevents the rippled glass template, and consequently the nanoantennas array confined on it, from showing grating optical effects which would lead to a more complex engineering of the color routing properties of our self-organized large area platform.

**NUMERICAL OPTICAL MODEL**

For the numerical analysis of the nanostrip antennas we employed a commercial software (Comsol Multiphysics 5.3), implementing the full-vectorial finite element method for scattered field formalism in two dimensions. We assumed a circular computational domain with 500 nm radius, surrounded by perfectly matched layers (PML) with scattering boundary conditions. The effective environment approximation was assumed (in accord with, e.g., Refs. 1,2) by embedding the nanostrips into a homogeneous dielectric medium with non-dispersive and lossless permittivity.

The sketch of the bimetallic nanoantenna is shown in Fig. S2a. To avoid numerical artifacts, the vertices of the Au and Ag nanostrips have been rounded with 5 nm and 15 nm radius of curvature, respectively. The FEM mesh was accordingly defined so to resolve these radii with at least 5 elements. For the dielectric domain we set a maximum mesh element size corresponding to $\lambda/(5\sqrt{\varepsilon_{eff}})$, with $\lambda$ the optical wavelength in vacuum and $\varepsilon_{eff} = 2.05$ the effective permittivity of the dielectric. The latter was estimated according to the so-called *effective environment approximation* as the mean value between the permittivity of the silica substrate (2.13 and air), corrected with a further increase (of ~0.48, fitted by matching the spectral position of the plasmonic resonances) to take into account the conformal non stoichiometric silica spacer in-between the two nanostrips.

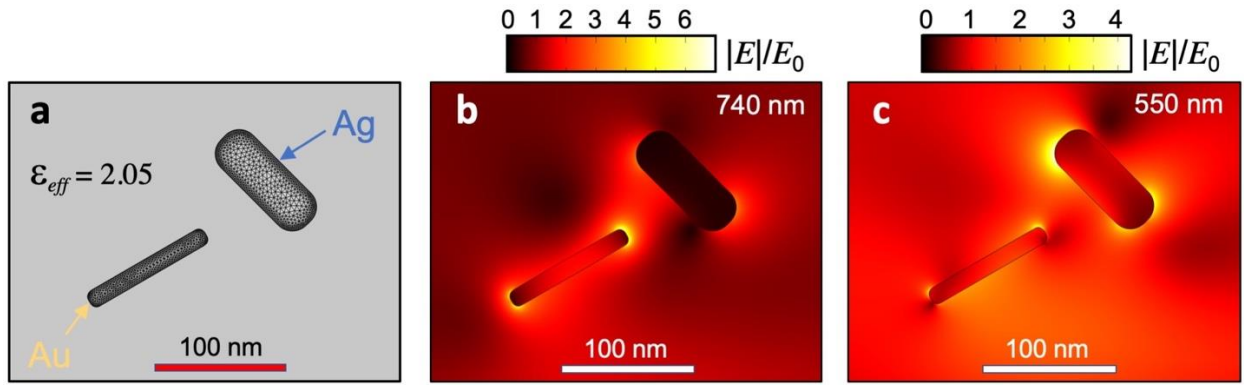

**Figure S2:** *(a) Sketch of the simulated 2D geometry (also showing FEM mesh elements, only in the nanowires for better reading). (b)-(c) Simulated near field enhancement at (b) 750 nm and (c) 550 nm, which approximately correspond, respectively, to the dip and peak wavelengths in the simulated directivity spectra in the bottom panel of Fig. 4c (see main text).*

Starting from the scattered electric $\mathbf{E}_S$ and magnetic $\mathbf{H}_S$ vector fields (numerically solved for as a function of $\lambda$), the total extinction cross-section is computed as $\sigma_E = \sigma_A + \sigma_S$, with $\sigma_A$ and $\sigma_S$ the total absorption and scattering cross-section spectra, respectively given by:

$$\sigma_A(\lambda) = \frac{\pi c \varepsilon_0 \varepsilon''_m(\lambda)}{\lambda I_0} \iint_{metal} |\mathbf{E}(\mathbf{r},\lambda)|^2 dS$$

$$\sigma_S(\lambda) = \frac{1}{2I_0} \int_\Sigma Real\{\mathbf{E}_S(\mathbf{r},\lambda) \times \mathbf{H}_S^*(\mathbf{r},\lambda)\} \cdot \mathbf{n} dl$$

In above formulas, $c$ is the speed of light in vacuum, $\varepsilon_0$ is the vacuum permittivity, $\Sigma$ is a circle surrounding the nanoscatterers, and $\mathbf{n}$ is the outward pointing normal vector of the circle. Note that since our numerical model is in 2D, the cross-sections above detailed are lengths and not areas, and the intensity $I_0$ of the incident plane wave, is thus measured in W/m.

For the metal permittivity spectrum $\varepsilon_m(\lambda) = \varepsilon'_m(\lambda) + i\,\varepsilon''_m(\lambda)$, we assumed the analytical model provided in Ref. 3 for Au, and in Ref. 4 for Ag, fitted on Johnson and Christy database. Note that, in order to mimick the inhomogenous broadening due to the dispersion of size (or periodicity) in our samples, we assumed an effective Drude damping parameter $\Gamma$ as large as 5 times the ideal value $\Gamma_0$ reported in Refs. 3,4. This is in agreement with other studies on poly-dispersed plasmonic systems (see e.g. Refs. 5,6).

With the total extinction cross-section at hand, the transmittance of the sample at normal incidence is estimated as following:

$$T(\lambda) = exp\left[-\frac{\sigma_E(\lambda)}{\eta L}\right]$$

where $L = 200$ nm is the measured average periodicity of the sample and $\eta$ a dimensionless fitting parameter of the order of 1 (in our simulations we set $\eta = 0.45$).

Concerning the far-field scattering patterns, we employed the FAR-FIELD procedure in Comsol, implementing the Stratton-Chu formulas (see, e.g., Ref. 7) using Σ as the aperture enclosing our 2D antennas (either being the Ag or Au monomer, or the Ag-Au dimer).

**OPTICAL CHARACTERIZATION**

VIS-NIR extinction measurements were performed at normal incidence using a halogen-deuterium compensated lamp (DH-2000-BAL, Mikropak) as source and a solid-state spectrometer (HR4000, Ocean Optics), operating in the wavelength range 300−1100 nm, as detector. Scattering spectra were acquired by employing a custom made scatterometer which can collect light as a continous function of the azimuthal angle, for a fixed selection of polar angles. The sample was illuminated from the bare flat glass side with a Visible and Near-Infra-Red broadband laser source (SuperK COMPACT by NKT Photonics) pulsed at high frequency (~10 kHz) in the super-continuum regime, to have a stronger signal intensity. A fiber coupled spectrometer Ocean Optics HR4000 was again used as detector. For both extinction and scattering measurements, the light source was linearly polarized to investigate the anisotropic optical properties of the nanoantennas array.

In Fig. S3 we show digital camera images of a fragment of a complete Au-Ag dimer covered sample, looked upon from the Ag side (Fig. S3a) and Au side (Fig.S3b) when illuminated from below by a white lamp. The two digital pictures where acquired with the same aperture and exposure parameters. Fig. S3 clearly shows the passive, macroscopic color routing action taking place over the whole sample.

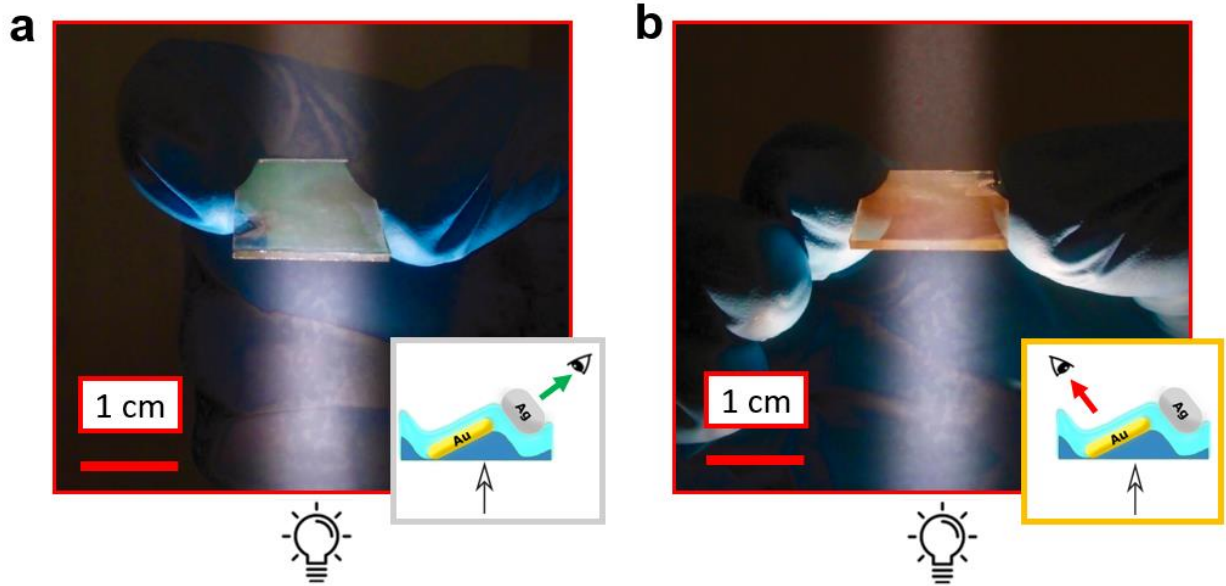

**Figure S3:** *Pictures of a Au-Ag dimer sample fragment illuminated from below by a white lamp, looked upon from the Ag side (a) and Au side using the same acquisition conditions (aperture F 3.5, exposure time 1/250 s), (b). The insets present a sketch of the picture acquisition geometry. The incoming white light has been artistically rendered for clarity.*

In Fig. S4 we show the measured directivity data for the Au, Ag and Au/Ag NSA reported in the main manuscript in Fig. 4a-b-c respectively, but using the same dB scale for all the panels. Fig. S4 evidences how the signal-to-noise ratio is similar for all the three different considered configurations when the directivity is not steeply changing with wavelenght.

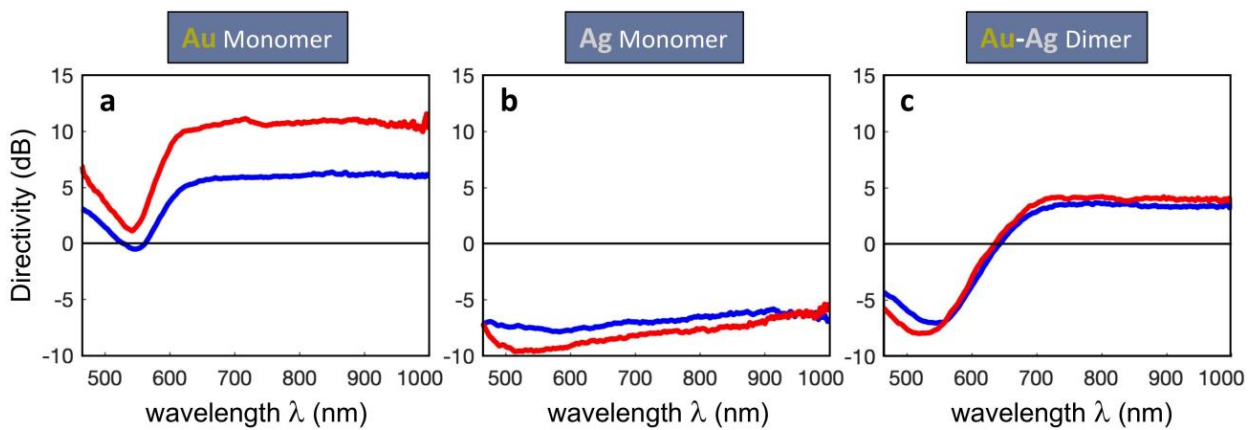

**Figure S4:** *Measured (a, b, c) directivities at two different polar angles (Θ = 30° in blue, Θ = 50° in red) for: (a) Au NSA monomer, (b) Ag NSA monomer and (c) Au-Ag NSA dimer configurations, respectively.*